\documentclass[hyper]{JHEP} 

\usepackage{epsfig}

\newcommand\fverb{\setbox\pippobox=\hbox\bgroup\verb}

\newcommand\fverbdo{\egroup\medskip\noindent%

            \fbox{\unhbox\pippobox}\ }

\newcommand\fverbit{\egroup\item[\fbox{\unhbox\pippobox}]}

\newbox\pippobox

\title{String in Ho\v{r}ava-Lifshitz Gravity}
\author{J. Kluso\v{n}\\
Department of
Theoretical Physics and Astrophysics\\
Faculty of Science, Masaryk University\\
Kotl\'{a}\v{r}sk\'{a} 2, 611 37, Brno\\
Czech Republic\\
E-mail: \email{klu@physics.muni.cz}}
\preprint{\hepth{1002.2849}}

 \abstract{We generalize recent analysis of the
 dynamics of point particle
 in Ho\v{r}ava-Lifshitz background
to the case of string probe when  we replace the
Hamiltonian constraint of the Polyakov string
with the constraint that breaks Lorentz invariance
of target space-time. Then we find corresponding Lagrangian
and argue that the world-sheet theory is invariant
under foliation preserving diffeomorphism. Finally
we discuss the Hamiltonian dynamics and show that
this is well defined on condition that the world-sheet
lapse function obeys the projectability condition.}
\keywords{Bosonic String, Ho\v{r}ava-Lifshitz Background}

\def\mH{\mathcal{H}}

\def\partt{\partial_\tau}
\def\parts{\partial_\sigma}

\def\bx{\mathbf{x}}

\newcommand{\bT}{\mathbf{T}}

\newcommand{\mL}{\mathcal{L}}

\def\pb #1{\left\{#1\right\}}

\begin{document}
\section{Introduction and Summary}\label{first}
Recently Petr Ho\v{r}ava proposed new
intriguing approach for the formulation
of UV finite  quantum theory of gravity
\cite{Horava:2009uw,Horava:2008ih,Horava:2008jf}
\footnote{This conjecture was then
studied in several papers, see for
example
\cite{Koyama:2009hc,Park:2009gf,Chen:2009vu,Wang:2009azb,
Tang:2009bu,Boehmer:2009yz,Leon:2009rc,Blas:2009qj,Wu:2009ah,Iorio:2009qx,
Ding:2009pq,Carloni:2009jc,Lee:2009rm,Kluson:2009rk,Wang:2009yz,Peng:2009uh,
Mukohyama:2009tp,Blas:2009yd,Germani:2009yt,Myung:2009sa,Park:2009zra,
Sakamoto:2009ww,Nojiri:2009th,Calcagni:2009qw,Mukohyama:2009mz,Gao:2009ht,
Kim:2009zn,Sotiriou:2009bx,Li:2009bg,Cai:2009dx,Mukohyama:2009zs,Sotiriou:2009gy,
Gao:2009bx,Piao:2009ax,Cai:2009ar,Nastase:2009nk,Brandenberger:2009yt,
Mukohyama:2009gg,Lu:2009em,Kluson:2009sm,Kiritsis:2009sh,Calcagni:2009ar,
Takahashi:2009wc,Visser:2009fg,Kocharyan:2009te,Kiritsis:2009rx,
Chen:2009jr,Saridakis:2009bv,Bogdanos:2009uj,Cai:2009in,Chakrabarti:2009ku,
Jacobson:2010mx,Chaichian:2010yi,Blas:2009ck,Henneaux:2009zb,
Kiritsis:2009vz,Papazoglou:2009fj,Kluson:2009xx,
Wang:2009rw,Cai:2010ud,Greenwald:2009kp,
Ghodsi:2009zi,Ghodsi:2009rv,Majhi:2009xh,Konoplya:2009ig,Colgain:2009fe,
Cai:2009pe,Cai:2009qs,Cai:2009ph}.}.
The basic idea of this theory is to
modify the UV behavior of the general
theory so that the theory is
perturbatively renormalizable. However
this modification is only possible on
condition when we abandon
 Lorentz symmetry in the high energy regime: in
this context, the Lorentz symmetry is
regarded as an approximate symmetry
observed only at low energy.

It is important to stress that there
are two classes of Ho\v{r}ava-Lifshitz
theories. One is the class of the
projectable theories that are
characterized by property that the
lapse function is restricted to depend
on time only. The other one  is the
non-projectable class where the lapse
is allowed to depend on both space and
time. Very nice discussion of the
consistency of the non-projectable
Ho\v{r}ava-Lifshitz theory was
performed in \cite{Henneaux:2009zb}
that suggests inconsistency of given
theory. For that reason it seems that
the projectable version of
Ho\v{r}ava-Lifshitz theory is the only
consistent one.

The results derived during the analysis
of Ho\v{r}ava-Lifshitz gravities
imply that it is an
interesting problem to study properties
of theories  with broken  general
covariance. One such an interesting example
of Lorentz breaking theory is the
Hamiltonian formulation
of the point particle in
Ho\v{r}ava-Lifshitz theory that was recently
analyzed  in
\cite{Capasso:2009fh,Suyama:2009vy,Romero:2009qs,Mosaffa:2010ym,
Kiritsis:2009rx,Rama:2009xc}.
The basic idea of this
Lorentz breaking Hamiltonian
formalism is that the time and spatial
components of momenta are treated
differently with very interesting
consequences for the notion of causality
and   formation
of black holes.

As the next logical step in this direction
is to try
to implement this approach for
the construction of extended objects,
as for example string. The goal of
this paper is to proceed in this direction
and construct new string theories
that we  call as Lorentz-breaking string
theories (LBS). Explicitly, we  study
properties of two dimensional gravity
coupled to the scalar field when the
Hamiltonian constraint breaks the
target space Lorentz invariance and
analyze its consequence for dynamics of
given string. As opposite to the
relativistic case when the
step from the point particle action
 to string action is straightforward
in case it turns out that in case of
LBS there are several ways how to do
it. We begin with the Polyakov action
and find its Hamiltonian formulation.
Then we replace the  Hamiltonian
constraint with its Lorentz breaking
generalization. Since now the
world-sheet modes are spatial dependent
we find more possibilities how to
construct such a Hamiltonian constraint
as opposite to the case of the point
particle in Ho\v{r}ava-Lifshitz
background. The characteristic property
of the classical Polyakov action is
that the two dimensional gravity is non
dynamical. On the other hand  we will
argue that for the consistency of the
LBS theory it is necessary that the
spatial component of the world-sheet
metric becomes dynamical.  Further, we
also show that the resulting
world-sheet theory is not invariant
under full two dimensional
diffeomorphism but only under the
world-sheet foliation preserving one.
Finally, the consistency of the
Hamiltonian dynamics of LBS theory
again implies that the world-sheet
lapse has to obey the projectability
condition which is similar result as in
case of Ho\v{r}ava-Lifshitz gravity.

Let us conclude our results. We
construct new form of string theories
 with broken target space
Lorentz invariance.
 We should however  stress
that these string theories have to
be  considered as  toy models for the
 study of the systems with breaking
 symmetries.
For example,  since LBS
 theories are manifestly non-linear
 even in flat space-time it is not
 clear how to formulate their quantum
 version. In fact, since it is not
 known whether Ho\v{r}ava-Lifshitz
 gravity can be formulated in the
 framework of string theory it is also
 questionable how to formulate the
 action for string probe in given
 background.
However despite of these
 doubts we believe that the study of
 such system is interesting in its own
 and should be extended in may ways.
 In particular, we  would like to see how
these LBS theories  behave under target
space T-duality transformations. We
would like also perform similar
analysis in case of higher dimensional
extended systems, as  Dp-branes. We
hope to return to these problems in
future.

The organization of this paper is as
follows. In the next section
(\ref{second}) we review the
Hamiltonian and Lagrangian formulation
of the probe particle in
Ho\v{r}ava-Lifshitz background. Then in
section (\ref{third}) we study the
string probe in given background and
formulate LBS theories. In section
(\ref{fourth}) we study their
Hamiltonian formalism and calculate the
algebra of constraints.

\section{Dynamics of point particle
in Ho\v{r}ava-Lifshitz
gravity}\label{second}
 In this section
we review the Hamiltonian dynamics of
the particle in Ho\v{r}ava-Lifshitz
gravity. This section can be considered
as  the review and extension of
analysis performed in
\cite{Capasso:2009fh,Suyama:2009vy,Romero:2009qs,Mosaffa:2010ym,
Kiritsis:2009rx,Rama:2009xc}.

Let us consider space-time
$\mathcal{M}$ labeled with coordinates
$t,\bx=(x^1,\dots,x^D)$ with the metric
in ADM form
\begin{equation}
g_{00}=-N^2+N_ih^{ij}N_j \ , \quad
g_{0i}=N_i \ , \quad g_{ij}=h_{ij} \ ,
\quad  \det g=-N^2\det h \
\end{equation}
with inverse
\begin{equation}
g^{00}=-\frac{1}{N^2} \ , \quad
g^{0i}=\frac{N^i}{N} \ , \quad
g^{ij}=h^{ij}-\frac{N^i N^j}{N^2} \ .
\end{equation}
Now we are ready to formulate  the Hamiltonian
description of   point
particle in the background with
broken Lorenz invariance.
 This problem was studied
in \cite{Capasso:2009fh,Suyama:2009vy,Romero:2009qs,Mosaffa:2010ym,
Kiritsis:2009rx,Rama:2009xc}
where it was argued that
 Hamiltonian constraint should have the form
\begin{equation}\label{HT}
H_T=-\frac{1}{N^2}(p_t-N^i p_i)^2+h(p^2)\approx 0 \ , \quad
p^2=p_ip_j h^{ij} \
\end{equation}
so that the Hamiltonian is
given as
\begin{equation}\label{HT1}
H=\lambda(t)H_T \ ,
\end{equation}
where $\lambda(t)$ is the Lagrange
multiplier that express the fact that
the Hamiltonian is proportional to the
first class constraint (\ref{HT}). This
form of   Hamiltonian constraint
reflects the non-relativistic nature of
the Ho\v{r}ava-Lifshitz gravities where
generally $h(A)=A^n$ for some positive
$n$. In particular, the analysis of the
dynamics of the probe particle that is
governed by this Hamiltonian brought
many interesting results. For example,
it was argued that particles with such
a form of the Hamiltonian
does not feel the presence of the
horizon in the background of black hole
\footnote{For further details we
recommend the original papers
\cite{Capasso:2009fh,Suyama:2009vy,Romero:2009qs,Mosaffa:2010ym,
Kiritsis:2009rx,Rama:2009xc}.}.

Now we  determine Lagrangian
corresponding to the Hamiltonian
(\ref{HT1}).
It is useful to  introduce two
non-dynamical  modes $A,B$ together
with their conjugate momenta $P_A,P_B$
and rewrite the Hamiltonian constraint
in the form
\begin{equation}\label{HTex}
H_T=-\frac{1}{N^2}(p_t-N^i
p_i)^2+B(p^2-A) +h(A) \
\end{equation}
so that the Hamiltonian takes the form
\begin{equation}
H=\lambda(\tau)H_T+v_AP_A+v_BP_B \ ,
\end{equation}
 where $v_A,v_B$
are Lagrange multipliers that ensure
that $P_A,P_B$ are  primary
constraints of the theory
\begin{equation}
P_A\approx 0 \ , \quad P_B \approx 0 \ .
\end{equation}
In fact, the consistency of these constraints
with the time evolution of the system
 implies two secondary
constraints $G_A,G_B$. Explicitly
\begin{equation}
\frac{dP_A}{d\tau}=
\pb{P_A,H}=\lambda(\tau)(B-h'(A))=\lambda(\tau)G_A\approx
0 \ , \quad  h'(A)\equiv \frac{dh(A)}{dA}
\end{equation}
and
\begin{equation}
\frac{dP_B}{d\tau}= \pb{P_B,H}=
-\lambda(\tau)(p^2-A)=\lambda(\tau)G_B
\approx 0  \ .
\end{equation}
 It can be shown that the collection
of the constraints $(P_A,P_B,G_A,G_B)$
form the set of the second class
constraints that can be explicitly
solved as $p^2=A$ and $B=h'(A)$.
Then inserting these results
into (\ref{HTex}) we recovery the
original Hamiltonian constraint
(\ref{HT}).

Now with the help of
 the extended Hamiltonian it is
easy to find the corresponding Lagrangian since
\begin{eqnarray}
\frac{dt}{d\tau}&=&
\pb{t,H}=-\frac{2\lambda}{N^2}
(p_t-N^i p_i) \ ,
\nonumber \\
\frac{dx^i}{d\tau}&=&
\pb{x^i,H}=\frac{2\lambda N^i}{N^2}
(p_t-N^jp_j)+2\lambda B h^{ij}p_j \ ,
\nonumber \\
\frac{d A}{d\tau}&=&
\pb{A,H}=v_A \ , \quad
\frac{dB}{d\tau}=\pb{B,H}=
v_B \ .
\nonumber \\
\end{eqnarray}
Then it is easy to
find corresponding Lagrangian
\begin{eqnarray}\label{Lpar}
L&=&p_M \frac{dx^M}{d\tau}+P_A\frac{dA}{d\tau}+
P_B\frac{dB}{d\tau}
-H=
\nonumber \\
&=&-\frac{N^2}{4\lambda }\left(\frac{dt}{d\tau}\right)^2+
\frac{1}{4B\lambda }V^i V^j h_{ij}+\lambda BA- \lambda h(A) \ ,
\nonumber \\
\end{eqnarray}
where
\begin{equation}
V^i=\frac{dx^i}{d\tau}-N^i \frac{dt}{d\tau} \ .
\nonumber \\
\end{equation}
As the final step we integrate out $A,B$ from
(\ref{Lpar}). Explicitly,
the variation of (\ref{Lpar}) with respect
to $A$ and $B$ implies the equation of
motion
\begin{eqnarray}\label{eqBV}
& & B-h'(A)=0 \ , \nonumber \\
& &-\frac{1}{4B^2\lambda }V^i V^j h_{ij}+A \lambda=0
\ .
\nonumber \\
\end{eqnarray}
The first equation in (\ref{eqBV}) implies
 $B=h'(A)$. Inserting
this result into the  second equation
 in (\ref{eqBV})
 we obtain an algebraic
equation for $A$
\begin{equation}\label{eqA}
-\frac{1}{4 h'^2(A)\lambda^2}V^i V^j h_{ij}+A=0
 \ .
 \end{equation}
 For given  function $h(A)$ this equation can be
 solved for $A$
 \begin{equation}\label{solA}
A=\Psi \left(\frac{1}{4\lambda^2}
V^i V^j h_{ij}\right) \ .  \nonumber \\
\end{equation}
Then inserting  (\ref{eqBV}), (\ref{eqA}) and
(\ref{solA}) into (\ref{Lpar})
 we find the Lagrangian
in the form
\begin{equation}\label{L}
L=\lambda\left[-\frac{N^2}{4\lambda^2}
\left(\frac{dt}{d\tau}\right)^2
+\frac{2}{h'\left(\Psi(\frac{1}{4\lambda^2}V^i V^j h_{ij}) \right)}\frac{1}{4\lambda^2}
V^i V^j h_{ij}
- h\left(\Psi\left(\frac{1}{4\lambda^2}
V^i V^j h_{ij} \right)\right)\right] \ .
\end{equation}
Now we show that the action $S=\int
d\tau L$ with $L$ given in (\ref{L}) is
invariant under the world-line
diffeomorphism
\begin{equation}\label{wld}
\tau'=\tau'(\tau) \ .
\end{equation}
As the first step we postulate
 that
 $\lambda(\tau)$ transforms under
 (\ref{wld}) as
\begin{equation}
\lambda'(\tau')=\lambda(\tau)\frac{d\tau}{d\tau'}
\end{equation}
so that $d\tau \lambda(\tau)$ is
invariant under (\ref{wld}). Further we
demand  that  $t,x^i$ behave as scalars
 under (\ref{wld})
\begin{equation}
t'(\tau')=t(\tau) \ , \quad
x'^i(\tau')=x^i(\tau) \ .
\end{equation}
Consequently we find
\begin{equation}
V'^i(\tau')=
V^i(\tau)\frac{d\tau}{d\tau'} \ ,
\quad \frac{dt'(\tau')}{d\tau'}
=\frac{dt(\tau)}{d\tau}\frac{d\tau}{d\tau'} \ .
\end{equation}
Using these results we immediately see
that $\frac{1}{\lambda^2}V^i V^j
h_{ij}$ and
$\frac{1}{\lambda^2}\left(\frac{dt}{d\tau}\right)^2$
are invariant under (\ref{wld}). In summary
we see that the action
is invariant under (\ref{wld}).

 We conclude  this section we
some examples of the function $h=h(A)$.
As the first one we consider
$h=\frac{1}{2n}A^{n}$. For this
function the equation (\ref{eqA}) has
the solution
\begin{equation}
A=\left(\frac{1}{\lambda^2}V^i V^j h_{ij}\right)^{\frac{1}{2n-1}}
\end{equation}
and consequently the Lagrangian takes
the form
\begin{equation}
L=\lambda\left[-\frac{N^2}{4\lambda^2}
\left(\frac{dt}{d\tau}\right)^2
+(1-\frac{1}{2n})\left(\frac{1}{\lambda^2}
Vî V^j h_{ij}\right)^{
\frac{n}{2n-1}}\right] \ .
\end{equation}
This is the Lagrangian studied in
\cite{Capasso:2009fh,Suyama:2009vy,Romero:2009qs,Mosaffa:2010ym,
Kiritsis:2009rx,Rama:2009xc}.

As the second example we consider the
function $h=\sqrt{1+A}$. For this
function  the equation (\ref{eqA}) has
the solution
\begin{equation}
A=\frac{V^iV^jh_{ij}}{\lambda^2(1-\frac{1}{\lambda^2}
V^iV^jh_{ij})} \
\end{equation}
and consequently the Lagrangian takes the form
\begin{equation}
L=\lambda \left[-\frac{N^2}{4\lambda^2}
\left(\frac{dt}{d\tau}\right)^2
-\sqrt{1-\frac{1}{\lambda^2} V^iV^j
h_{ij}}\right] \ .
\end{equation}

\section{Lorentz Breaking String
Theory} \label{third} In this section
we generalize the analysis presented in
the previous section to  the case of
string probe in Ho\v{r}ava-Lifshitz
background when we study properties of
the string with the new form of the
Hamiltonian constraint that generalizes
the point particle Hamiltonian
constraint studied in the previous
section. In fact, since  it is not know
how or whether it is possible to
formulate Ho\v{r}ava-Lifshitz gravities
in the framework of string theories we
mean that it is legitime to consider
string probe in given background with
 Hamiltonian constraint that breaks
 the Lorentz invariance of the target
 space-time which is the characteristic
 property of Ho\v{r}ava-Lifshitz background.

As in case of point particle
LBS theory is defined using the Hamiltonian
formalism that is based on the Hamiltonian
analysis of relativistic string.
 For that reason we now review
the Hamiltonian formulation of the
Polyakov string action in general
background.
\subsection{ Hamiltonian
formulation of Polyakov Action}
 Let us consider the
Polyakov action
 in
general background
\begin{equation}\label{Polact}
S=-\frac{1}{4\pi\alpha'} \int d^2\sigma
\sqrt{-\gamma}\gamma^{\alpha\beta}
g_{MN}\partial_\alpha x^M\partial_\beta
x^N  \ ,
\end{equation}
where $\gamma_{\alpha\beta}$ is a
two-dimensional world-sheet metric and
$ \sigma^\alpha \ , \alpha,\beta=0,1 \
,
 \sigma^0=\tau \ ,
\sigma^1=\sigma$ are
world-sheet coordinates.
 Further, $x^M\ ,
M,N=0,\dots,D$ are modes that
parameterize the embedding of the
string into target space-time with
the background metric $g_{MN}$.

In order to formulate the Hamiltonian
formalism from  the action
(\ref{Polact}) it is convenient to use
$1+1$ formalism for
 the world-sheet
metric
\begin{equation}
\gamma_{\alpha\beta}= \left(
\begin{array}{cc}
-n^2_\tau+ \frac{1}{\omega
}n_\sigma^2 &
n_\sigma \\
 n_\sigma & \omega \\
\end{array}\right) \ ,
\end{equation}
where $n_\tau$ is world-sheet lapse,
$n_\sigma$ is world-sheet shift and
$\omega$ is spatial part of world-sheet metric.
Then it is easy to see that
\begin{equation}
\det \gamma=-n_\tau^2 \omega \ , \quad
\gamma^{\alpha\beta}=
\left(\begin{array}{cc}
-\frac{1}{n^2_\tau}
 & \frac{n^\sigma}{ n^2_\tau} \\
 \frac{n^\sigma}{ n^2_\tau}
 &
 \frac{1}{\omega}-
 \frac{n^\sigma n^\sigma}{n_\tau^2  }
 \\ \end{array}\right) \ ,
\end{equation}
where we defined
\begin{equation}
n^\sigma\equiv \frac{n_\sigma}{\omega}
\ .
\end{equation}
With this form of the world-sheet
metric the action (\ref{Polact}) takes
the form
\begin{eqnarray}\label{Polact2}
S&=&\frac{1}{4\pi\alpha'} \int
d^2\sigma n_\tau\sqrt{\omega}
(g_{MN}\nabla_\tau x^M\nabla_\tau
x^N-\frac{1}{\omega}
g_{MN}\partial_\sigma x^M\partial_\sigma x^N) \ , \nonumber \\
 \nonumber \\
\end{eqnarray}
where
\begin{equation}
\nabla_\tau x^M=\frac{1}{n_\tau}
(\partial_\tau x^M-
n^\sigma\partial_\sigma x^M) \ .
\end{equation}
We  introduce the momenta
$\pi^\tau,\pi^\sigma,\pi^\omega$
conjugate to $n_\tau,n_\sigma$ and $\omega$
that have non-zero  Poisson brackets
\begin{equation}
\pb{n_\tau(\sigma),\pi^\tau(\sigma')}=
\delta(\sigma-\sigma') \ , \quad
\pb{n_\sigma(\sigma),\pi^\sigma(\sigma')}=
\delta(\sigma-\sigma') \ , \quad
\pb{\omega(\sigma),
\pi^\omega(\sigma')}=\delta(\sigma-
\sigma') \ .
\end{equation}
Since there are no time derivatives of
world-sheet metric $\gamma$ in
(\ref{Polact2}) the conjugate momenta
are primary constraints of theory
\begin{equation}\label{pricon}
\pi^\tau=\frac{\delta S}{\delta \partt
n_\tau}\approx 0 \ , \quad
\pi^\sigma=\frac{\delta S}{\delta
\partt n_\sigma}\approx 0 \ , \quad
\pi^\omega=\frac{\delta S}{\delta
\partt \omega}\approx 0 \ .
\end{equation}
As the next step we introduce the
momenta $p_M$ conjugate to $x^M$
\begin{equation}
p_M=\frac{\delta S}{\delta
\partt x^M}=\frac{1}{2\pi\alpha'} \sqrt{\omega}
g_{MN}\nabla_\tau x^N \
\end{equation}
with standard Poisson brackets
\begin{equation}
\pb{x^M(\sigma),p_N(\sigma')}=
\delta^M_N\delta(\sigma-\sigma') \ .
\end{equation}
Then it is easy to find  the
Hamiltonian density
\begin{eqnarray}\label{H}
\mH=\partt x^Np_M-\mL= n_\tau \mH_\tau+
n^\sigma
 \mH_\sigma
\ , \nonumber \\
\end{eqnarray}
where
\begin{eqnarray}\label{mHTS}
 \mH_T&=&\frac{1}{4\pi\alpha'\sqrt{\omega}}
\left((2\pi\alpha')^2p_M
g^{MN} p_N+
\parts x^M
g_{MN}\parts x^N\right) \ , \nonumber \\
\mH_S&=&\parts x^Np_N \ . \nonumber \\
\end{eqnarray}
Using the  Hamiltonian (\ref{H})
we easily determine
the time evolution of the primary
constraints as
\begin{eqnarray}
\partt \pi_\tau&=&\pb{\pi_\tau,H}=
-\mH_\tau \ ,
\nonumber \\
\partt \pi_\omega&=&\pb{\pi_\omega,H}=
\frac{n_\tau}{2\omega}\mH_\tau+
\frac{1}{\omega^2}n_\sigma \mH_\sigma
 \ ,
\nonumber \\
\partt \pi_\sigma&=&\pb{\pi_\sigma,H}=
-\mH_\sigma  \ .  \nonumber \\
\end{eqnarray}
Since the  constraints (\ref{pricon})
 have to be
preserved during the time evolution of
the system the equations above imply an
existence of the  secondary constraints
\begin{equation}\label{secondco}
\mH_\tau\approx 0 \ , \quad \mH_\sigma\approx 0
\ .
\end{equation}
The existence of these constraints will
be the starting point for the
construction of Lorentz breaking string
theories.
\subsection{Lorentz Breaking String Action}
The construction of Lorentz breaking
string theories  is based on the
generalization of the point particle
analysis reviewed in the first section.
Due to the fact that the world-sheet
modes depend on $\sigma$ there are
several possibilities how to do it. We
consider the case  when we presume that
$p_t$ and $t$ are functions of $\tau$
only and suggest following
form  of
constraints (\ref{mHTS})
\begin{eqnarray}
\mH_\tau&=&
-\frac{\pi\alpha'}{N^2\sqrt{\omega}}
(p_t-N^ip_i)^2 + \sqrt{\omega}
F\left(\frac{1}{\omega}
\left[\pi\alpha' p_i
h^{ij}p_j+\frac{1}{4\pi\alpha'}
\partial_\sigma x^i\partial_\sigma x^j
h_{ij}\right]\right) \ , \nonumber \\
\mH_\sigma&=& p_i\parts x^i \ , \nonumber \\
\end{eqnarray}
where $F$ is an  arbitrary function.
Our goal is to study  properties of
string theory with this form of the
Hamiltonian constraint $\mH_\tau$.

As the first step we  find the
Lagrangian formulation of given theory.
Following analysis performed in
 section (\ref{second}) we introduce
two non-dynamical fields $A,B$ in order
to rewrite the Hamiltonian constraint
into the form
\begin{eqnarray}\label{mHTAB}
\mH_\tau
&=&-\frac{\pi\alpha'}{N^2\sqrt{\omega}}
(p_t-N^ip_i)^2+
\nonumber \\
&+& B\left(\frac{1}{\omega
}\left[\pi\alpha' p_i
h^{ij}p_j+\frac{1}{4\pi\alpha'}
\partial_\sigma x^i\partial_\sigma x^j
h_{ij}\right]-A\right)+\sqrt{\omega}F(A)+v_A
P_A+v_B P_B \ ,
 \nonumber \\
\end{eqnarray}
where $v_A,v_B$ are Lagrange
multipliers that ensure that the
momenta $P_A,P_B$ vanish.  Then from
the Hamiltonian
\begin{equation}
H=\int d\sigma (n_\tau
\mH_\tau+n^\sigma
\mH_\sigma) \ , \
\end{equation}
where $\mH_\tau$ and $\mH_\sigma$  are
 given in (\ref{mHTAB})
we derive following equations of motion
for $t,x^i,A,B$
\begin{eqnarray}
\partt t&=&\pb{t,H}=
-\frac{2\pi\alpha'}{N^2\sqrt{\omega}}
n_\tau(p_t-N^i p_i) \ , \nonumber \\
\partt x^i&=&\pb{x^i,H}=
2\pi\alpha'\frac{n_\tau}{\sqrt{\omega}}
\frac{N^i}{N^2} (p_t-N^j
p_j)+2\pi\alpha'\frac{n_\tau}{\omega}
B h^{ij}p_j+\frac{1}{\omega}n_\sigma
\parts x^i \ ,
\nonumber \\
\partt A&=&\pb{A,H}= n_\tau v_A \ , \quad  \partt B=\pb{B,H}=
n_\tau v_B \ . \nonumber \\
\end{eqnarray}
Then it is easy to find corresponding
Lagrangian density
\begin{eqnarray}
L&=&\partt x^M p_M+\partt A P_A+\partt
B P_B-H=
\nonumber \\
&=&-\frac{N^2\sqrt{\omega}n_\tau}{4\pi\alpha'n^2_\tau}
(\partt t)^2+ \frac{n_\tau
\omega}{4\pi\alpha' } \frac{1}{B
n_\tau^2} V^i h_{ij}V^j-\nonumber
\\
&-&n_\tau B
 \left(\frac{1}{4\pi\alpha'\omega}
 \parts x^i \parts x^j h_{ij}-A\right)-n_\tau \sqrt{\omega}
F(A) \ ,
\nonumber \\
\end{eqnarray}
where
\begin{eqnarray}
V^i\equiv \partt x^i+\partt t
N^i-n^\sigma
\parts x^i \ .  \nonumber \\
\end{eqnarray}
Finally we integrate out non-dynamical
fields  $A$ and $B$ by solving their
equations of motion. The  equation of
motion for $A$ implies
\begin{equation}\label{eqB1}
B-\sqrt{\omega}F'(A)=0
\end{equation}
while the equation of motion for  $B$
gives
\begin{equation}\label{eqBA}
-\frac{1}{4\pi\alpha'}
\frac{\omega}{n^2_\tau B^2} V^i
h_{ij}V^j -n_\tau
\left(\frac{1}{4\pi\alpha\omega}
 \parts x^i \parts x^j h_{ij}-A\right)
=0 \ .
\end{equation}
For known $F(A)$ these equations can
be solved for $A$
\begin{equation}
A=\Psi\left(
\frac{1}{4\pi\alpha'}
\frac{1}{n^2_\tau}
V^i h_{ij}V^j,
\frac{1}{4\pi\alpha'
\omega}\parts x^i \parts x^j h_{ij}
\right) \ .
\end{equation}
As a result we  find that the Lagrangian
density in the form
\begin{equation}\label{mLfin}
\mL=\sqrt{\omega}n_\tau
\left[-\frac{N^2}{4\pi\alpha'n^2_\tau}
(\partt t)^2+ \frac{1}{2\pi\alpha'}
\frac{1}{ F'(\Psi) n_\tau^2} V^i
h_{ij}V^j- F(\Psi)\right] \ .
\nonumber \\
\end{equation}
Clearly this form of the Lagrangian is
completely different from the Polyakov
Lagrangian. It is very interesting to
study the consequence of the breaking
of the target space  covariance. In
particular we would like to identify
symmetries of the Lagrangian density
(\ref{mLfin}).
\subsection{Symmetries of the action}
In this section we study the symmetries
of the action $S=\int d\tau d\sigma
\mL$ where $\mL$ is given in
(\ref{mLfin}). We begin with the
discussion of the global symmetries of
the action from the point of view of
the string world-sheet theory. These
transformations correspond to the target
space foliation preserving diffeomorphism
\cite{Horava:2008ih}
\begin{eqnarray}\label{fpdG}
t'(\tau)&=&t(\tau)+f(t(\tau)) \ ,
\nonumber \\
x'^i(\tau,\sigma)&=&
x^i(\tau,\sigma)+\xi^i(t(\tau),x^i(\tau,\sigma))
\
\nonumber \\
\end{eqnarray}
so that
\begin{eqnarray}
\partt t'&=&
\partt t+\dot{f}
\partt t \ ,
\nonumber \\
\partt x'^i&=&
\partt x^i+
\partial_j \xi^i
\partt x^j+
\dot{\xi}^i \partt t
 \ , \nonumber \\
\parts x'^i&=&
\parts x^i+\partial_j
\xi^i
\parts x^j \ ,
\nonumber \\
\end{eqnarray}
where
\begin{equation}
\dot{F}=\frac{dF}{d t} \ , \quad
\partial_i F=\frac{\partial F}{\partial x^i} \
\end{equation}
for any $F$.
Note that under these transformations
the metric components transform as
\begin{eqnarray}
N'_i(t',\bx')&=& N_i(t,\bx) -N_i(t,\bx)
\dot{f}(t)-N_j(t,\bx)\partial_i
\zeta^j(t,\bx)-g_{ij}(t,\bx)
\dot{\zeta}^j(t,\bx) \ ,   \nonumber \\
N'(t')&=&N(t)-N(t) \dot{f}(t) \ , \nonumber \\
g'_{ij}(t',\bx')&=&g_{ij}(t,\bx)-
g_{il}(t,\bx)\partial_j
\zeta^l(t,\bx)-\partial_i
\zeta^k(t,\bx) g_{kj}(t,\bx) \ .  \nonumber \\
 \end{eqnarray}
Then it is easy to see that
\begin{eqnarray}
N'^2(t') (\partt t')^2&=&
N^2(t)(\partt t)^2  \ , \nonumber \\
(V^i h_{ij}V^j)'&=&
V^i
h_{ij}V^j \ ,
\nonumber \\
(\parts x^i h_{ij}\parts x^j)'&=&
\parts x^i h_{ij}\parts x^j \ , \nonumber \\
\end{eqnarray}
using also  the fact that $V^i$ transforms
under (\ref{fpdG}) as
\begin{eqnarray}
V'^i
=V^i+\partial_j
\xi^i V^j  \ .
\nonumber \\
\end{eqnarray}
Collecting all these results we
find that the Lagrangian density
(\ref{mLfin}) is invariant
under  (\ref{fpdG}).

Let us now discuss the local symmetries
of the action. We  argue that this
action is invariant under world-sheet
foliation preserving diffeomorphism
that are  generated by infinitesimal
transformations
\begin{equation}\label{fpd}
\delta \sigma \equiv
\sigma'-\sigma=\zeta(\tau,\sigma) \ , \quad
\delta \tau \equiv \tau'-\tau=f(\tau) \
.
\end{equation}
In the same way as in
\cite{Horava:2008ih} we find that
the world-sheet  metric components transform
under (\ref{fpd}) as
\begin{eqnarray}
n'_\sigma(\tau',\sigma')&=&
n_\sigma(\tau,\sigma)
-n_\sigma(\tau,\sigma)\parts
\epsilon(\tau,\sigma)
 -\partt f(\tau)
n_\sigma(\tau,\sigma)
 -\partt \epsilon(\tau,\sigma) \omega(\tau,\sigma)
\ ,
\nonumber \\
n'_\tau(\tau',\sigma')&=&
n_\tau(\tau,\sigma)-n_\tau(\tau,\sigma)
\partt f(\tau) \ ,
\nonumber \\
\omega'(\tau',\sigma')&=&
\omega(\tau,\sigma) -2\parts
\epsilon(\tau,\sigma)
\omega(\tau,\sigma) \ ,
\nonumber \\
n'^\sigma(\tau',\sigma')&\equiv&
\frac{n'_\sigma(\tau',\sigma')}{\omega'(\tau',\sigma')}=
n^\sigma(\tau,\sigma)
+n^\sigma(\tau,\sigma)
\parts \epsilon(\tau,\sigma)
-n^\sigma(\tau,\sigma)
\partt f(\tau)-
\partt \epsilon(\tau,\sigma) \ .
\nonumber \\
\end{eqnarray}
Then it is easy to see that
\begin{eqnarray}
d\tau' d\sigma' n'_\tau\sqrt{\omega'}
= d\tau d\sigma n_\tau
\sqrt{\omega} \ .
\nonumber \\
\end{eqnarray}
Clearly, the world-sheet
modes $x^M$ transform as
scalars under (\ref{fpd})
\begin{equation}
x'^M(\tau',\sigma')=
x^M(\tau,\sigma) \  .
\end{equation}
Then we easily show that
\begin{eqnarray}
\frac{1}{n'^2_\tau(\tau')}
(\partial_{\tau'}t'(\tau'))^2=
\frac{1}{n_\tau^2(\tau)} (\partt
t(\tau))^2
\end{eqnarray}
and
\begin{eqnarray}
(\frac{1}{n^2_\tau}V^i h_{ij}
V^j)'(\tau',\sigma')
=
\frac{1}{n^2_\tau}V^i h_{ij}V^j(\tau,\sigma)
\nonumber \\
\end{eqnarray}
using the fact that
\begin{eqnarray}
V'^i(\tau',\sigma')=
V^i(\tau,\sigma)-
V^i(\tau,\sigma)\partt f(\tau) \  .
\nonumber \\
\end{eqnarray}
In the same way we can show
that
\begin{eqnarray}
(\frac{1}{\omega}
\parts x^i \parts x^jh_{ij})'(\tau',\sigma')=
\frac{1}{\omega}
\parts x^i \parts x^jh_{ij}(\tau,\sigma) \ .
\nonumber \\
\end{eqnarray}
Again, collecting all these  results we
find that the action $S=\int d\tau
d\sigma \mL$ with $\mL$ given in
(\ref{mLfin}) is invariant under
world-sheet foliation preserving
diffeomorphism. In other words, the
Lorentz braking form of the Hamiltonian
constraint is consistent with the
world-sheet  theory where the full
diffeomorphism invariance is reduced.
It is important to  stress that at this
moment we still presume the general
dependence of $n_\tau$ on $\tau$ and
$\sigma$. In other words, we formulate
theory without projectability condition
imposed. We discuss the problem of
projectability in the next section.

Now we give explicit examples of the
function $F$. The first one is the
polynomial  function $F=\frac{1}{n}A^n$
known from the point particle analysis
in the Ho\v{r}ava-Lifshitz background.
 For this function
$B=\sqrt{\omega}A^{n-1}$ and then the
equation (\ref{eqBA}) leads to
 polynomial equation for $A$
where it is difficult to find their
roots analytically. Clearly the
resulting Lagrangian density is very
complicated and it is hardly to see how
such a form of the Lagrangian could be
deduced from the first principles.

As the second example we consider the
function $F(A)=\sqrt{1+A}$. For this
form of the function it is possible to
find solution (\ref{eqBA}). On the
other hand this function does not have
a support from the study of the probe
in Ho\v{r}ava-Lifshitz gravity.
Explicitly, the equation (\ref{eqB1})
implies
$B=\frac{1}{2\sqrt{1+A}}\sqrt{\omega} $
and then the equation (\ref{eqBA}) has
solution
\begin{eqnarray}
A=\frac{1}{1-\frac{4}{4\pi\alpha'}
\frac{V^i h_{ij}V^j}{n_\tau^2}}
\left(\frac{1}{4\pi\alpha'\omega}\parts
x^i h_{ij}
\parts x^j+
\frac{4}{4\pi\alpha'}\frac{1}{n^2_\tau}
V^i h_{ij}V^j\right) \ .
 \nonumber \\
\end{eqnarray}
Then it is easy to find the Lagrangian
density in the form
\begin{eqnarray}
\mL
=\sqrt{\omega}n_\tau\left[
-\frac{N^2}{4\pi\alpha'n^2_\tau}
(\partt t)^2-
\sqrt{\frac{1}{4\pi\alpha'\omega}
\parts x^i\parts x^j h_{ij}+\frac{1}{\pi\alpha'}
\frac{1}{n^2_\tau}
V^i h_{ij}V^j}\sqrt{1-\frac{1}{\pi\alpha'}
\frac{1}{n_\tau^2}  V^i h_{ij}V^j}\right] \ .
\nonumber \\
\end{eqnarray}
We conclude this section with
the second example of the
 Lorentz  breaking  Hamiltonian
constraint
\begin{eqnarray}
\mH_\tau=
-\frac{\pi\alpha'}{N^2\sqrt{\omega}} (p_t-N^ip_i)^2+
+ \sqrt{\omega} F\left(\frac{1}{\omega}
\pi\alpha' p_i h^{ij}p_j\right)+
\sqrt{\omega}G\left(\frac{1}{4\pi\alpha'\omega}
\partial_\sigma x^i h_{ij}\partial_\sigma x^j
\right) \ ,  \nonumber \\
\end{eqnarray}
where $F$ and $G$ are arbitrary
functions.  Then, following the
same analysis as above
we find the Lagrangian density
in the form
\begin{equation}\label{mLf2}
\mL=n_\tau\sqrt{\omega}
\left[-\frac{N^2}{4\pi\alpha'n^2_\tau}
(\partt t)^2+ \frac{1}{2\pi\alpha'}
\frac{1}{F'(\Psi) n_\tau^2} V^i
h_{ij}V^j- G\left(\frac{1}{4\pi\alpha'
\omega}\parts x^i\parts x^j
h_{ij}\right)- F(\Psi)\right] \ .
\nonumber \\
\end{equation}
Clearly this Lagrangian density is
invariant under the same set of
symmetries as (\ref{mLfin}) and hence
can be considered as another example of
Lorentz breaking string theory. Note
that for this form of the Hamiltonian
constraint it is possible to find
explicit form of the Lagrangian density
even in case of the function
 $F=\frac{1}{n}A^n$. In fact, it is
to see that $A$ is equal to
\begin{equation}
A=\left(\frac{n}{4\pi\alpha'}
\frac{V^i h_{ij}V^j}{n_\tau^2}\right)
^{\frac{1}{2n-1}}
\end{equation}
and consequently  the Lagrangian density takes the form
\begin{equation}
\mL=n_\tau\sqrt{\omega}
\left[-\frac{N^2}{4\pi\alpha'n^2_\tau}
(\partt t)^2+
n^{\frac{1-n}{2n-1}}
\left(\frac{1}{4\pi\alpha' n_\tau^2}
V^i h_{ij}V^j\right)^{\frac{n}{2n-1}}-
G\left(\frac{1}{4\pi\alpha'
\omega}\parts x^i\parts x^j
h_{ij}\right) \right] \ .
\nonumber \\
\end{equation}
This Lagrangian density takes similar
form as the  particle Lagrangian
studied in the second section. In other
words, the extension of the Lorentz
breaking Hamiltonian constraint from
the point particle to string is not
unique that should be compared with
relatively straightforward step from
relativistic particle to relativistic
string.
\section{Hamiltonian Formalism of Non-Relativistic
String Theory Revised}\label{fourth}
 In the previous
section we found  new class of
non-relativistic string theories when
we presumed  Hamiltonian constraint
that breaks Lorentz invariance of the
target space. However it is important
to stress that it is not possible to
change the Hamiltonian constraint
freely  without further  checking of
the consistency of given theory. In
fact, let us again consider the
Hamiltonian constraint
\begin{eqnarray}\label{mHtn}
\mH_\tau
=-\frac{\pi\alpha'}{N^2\sqrt{\omega}}
(p_t-N^ip_i)^2+  \sqrt{\omega}
F\left(\frac{1}{\omega}
\left[\pi\alpha' p_i
h^{ij}p_j+\frac{1}{4\pi\alpha'}
\partial_\sigma x^i\partial_\sigma x^j
h_{ij}\right]\right) \ . \nonumber \\
\end{eqnarray}
As follows from (\ref{mHtn}) the
Hamiltonian constraint depends on
$\omega$ in non-trivial way. However
this fact suggests that it is not
useful to impose the primary constraint
$\pi^\omega\approx 0$ since then the
requirement of the consistency of this
constraint with the time evolution of
the system  implies additional
constraints. Then in order to avoid
imposing additional constraint on the
system we demand
 that $\omega$ is a dynamical
mode with kinetic term in the action.
The simplest possibility is to add to
the string action  following kinetic
term
\begin{equation}\label{SKomega}
S^K_\omega= \frac{1}{2} \int d\tau
d\sigma n_\tau \sqrt{\omega} K_\sigma
\frac{1}{\omega^2}K_\sigma \ ,
\end{equation}
where
\begin{eqnarray}
K_\sigma&=&\frac{1}{n_\tau} (\partt
\omega-2\nabla_\sigma n_\sigma) \ ,
\nonumber \\
\nabla n_\sigma&=&\parts
n_\sigma-\Gamma n_\sigma\ , \quad
\Gamma=\frac{1}{2\omega}\parts \omega
 \ . \nonumber \\
 \end{eqnarray}
Note that $\Gamma$ transforms under the
world-sheet foliation preserving
diffeomorphism (\ref{fpd})  as
\begin{eqnarray}
\Gamma'(\tau',\sigma')=
\Gamma(\tau,\sigma)-\Gamma(\tau,\sigma)
\parts \epsilon(\tau,\sigma)-
\parts^2\epsilon(\tau,\sigma)
\nonumber \\
\end{eqnarray}
so that  $\nabla_\sigma n_\sigma$
transforms as
\begin{eqnarray}
(\nabla_\sigma n_\sigma)'
=\nabla_\sigma n_\sigma -2\nabla_\sigma
n_\sigma \parts \epsilon-\nabla_\sigma
n_\sigma
\partt f -\parts \partt \epsilon \omega
- \frac{1}{2}\partt \epsilon \parts
\omega \ . \nonumber \\
\end{eqnarray}
Then after some algebra we find that
$K_\sigma$ transforms as
\begin{eqnarray}
K_\sigma'(\tau',\sigma')=
K_\sigma(\tau,\sigma)-
2K_\sigma (\tau,\sigma)
\parts \epsilon(\tau,\sigma) \ .  \nonumber \\
\end{eqnarray}
This result implies that $S^K_\omega$
is manifestly invariant under
(\ref{fpd}). We should also stress that
this  choice of the kinetic term is the
minimal one. For example, we could in
principle include the term
$\frac{1}{\omega^2}K^2_\sigma $
 into the factor of
the function $F$. For simplicity we
consider the first option and we find
\begin{equation}
\pi^\omega=\frac{\delta S^K_\omega}
{\partt \omega}= \sqrt{\omega}n_\tau
\frac{1}{\omega^2n_\tau}K_\sigma \ .
\end{equation}
Then it is easy to see that  the total
Hamiltonian  takes the form
\begin{equation}
\mH=n_\tau \mH_\tau+ n^\sigma
\mH_\sigma+ \lambda_\tau
\pi^\tau+\lambda_\sigma \pi^\sigma \ ,
\end{equation}
where
\begin{eqnarray}
\mH_\tau&=&-\frac{\pi\alpha'}{N^2\sqrt{\omega}}
(p_t-N^ip_i)^2+\frac{1}{2\sqrt{\omega}}
\pi^\omega \omega^2 \pi^\omega +
\nonumber \\
&+& \sqrt{\omega}
F\left(\frac{1}{\omega}
\left[\pi\alpha' p_i
h^{ij}p_j+\frac{1}{4\pi\alpha'}
\partial_\sigma x^i\partial_\sigma x^j
h_{ij}\right]\right)
 \ , \nonumber \\
\mH_\sigma&=& p_i\parts x^i- 2\omega
\nabla_\sigma \pi^\sigma
\ . \nonumber \\
\end{eqnarray}
In other words, the LBS theory is
characterized by two primary
constraints $\pi^\tau\approx 0 \ ,
\pi^\sigma\approx 0$ together with two
secondary ones $\mH_\tau\approx 0 \ ,
\mH_\sigma\approx 0$ where  we still
presume that $n_\tau$ depends on $\tau$
and $\sigma$.

As the next step in the analysis of the
consistency of given theory we
calculate the algebra of constraints
and we   demand that it is closed.
Alternatively, we should check that the
consistency of  these constraints with
the time evolution of the system either
does not impose additional constraints
on the theory or they are not the
second class constraints which would
signal the pathological behavior of
given theory \cite{Henneaux:2009zb}.

In order to calculate the algebra of
constraints we introduce their smeared
form
\begin{eqnarray}
\bT_S(\xi)=\int d\sigma \xi(\tau,\sigma)
\mH_\sigma \ , \quad \bT_T(f)=
\int d\sigma f(\tau,\sigma)\mH_\tau \ . \nonumber \\
\nonumber \\
\end{eqnarray}
Then using the canonical Poisson
brackets
\begin{equation}
\pb{x^M(\sigma),p_N(\sigma')}=
\delta^M_N\delta(\sigma-\sigma') \ ,
\quad \pb{\omega(\sigma),
\pi^\omega(\sigma')}=\delta(\sigma-\sigma')
\end{equation}
we find
\begin{eqnarray}\label{bTSomega}
\pb{\bT_S(\xi),\omega(\sigma)}&=&
-\xi(\sigma)\parts \omega(\xi)-
2\parts \xi(\sigma) \omega(\sigma) \ ,
\nonumber \\
\pb{\bT_S(\xi), \pi^\omega(\sigma)}&=&
\parts \xi(\sigma) \pi^\omega(\sigma)
-\parts \pi^\omega (\sigma)\xi (\sigma)
\nonumber \\
\pb{\bT_S(\xi),p_i(\sigma)}&=&
-\parts \xi(\sigma)p_i(\sigma)-
\xi(\sigma)\parts p_i(\sigma) \ ,
\nonumber \\
 \pb{\bT_S(\xi),x^i(\sigma)}&=&
 -\parts x^i(\sigma)\xi(\sigma) \ . \nonumber \\
 \end{eqnarray}
Then we easily determine
\begin{eqnarray}
\pb{\bT_S(\xi),\mH_\sigma(\sigma)}=
-2\parts \xi(\sigma)\mH_\sigma(\sigma)-
\xi(\sigma) \parts \mH_\sigma(\sigma) \nonumber \\
\end{eqnarray}
and consequently
\begin{eqnarray}
\pb{\bT_S(\xi),\bT_S(\eta)}=
\int  d\sigma (\xi \parts \eta-\parts
\xi \eta) \mH_\sigma(\sigma)= \bT_S(\xi
\parts \eta-\parts \xi \eta) \ .
\nonumber \\
\end{eqnarray}
As the next step we determine
the Poisson bracket $\pb{\bT_S(\xi),
\bT_T(n)}$. Firstly, using  (\ref{bTSomega})
we find that
\begin{equation}
\pb{\bT_S(\xi),\mH_\tau(\sigma)}=
-\parts \xi(\sigma) \mH_\tau(\sigma)
-\xi(\sigma)
\parts \mH_\tau(\sigma)
\end{equation}
so that
\begin{eqnarray}
\pb{\bT_S(\xi),\bT_T(n)} =-\int d\sigma
 n(\sigma)(\parts \xi \mH_\tau+\xi
\parts \mH_\tau)=\bT_T(\xi\parts n) \ .
\nonumber \\
\end{eqnarray}
 As the final step we
calculate the Poisson bracket
$\pb{\bT_T(f),\bT_T(g)}$ where we
expect, with analogy with calculation
of Poisson bracket of Hamiltonian
constraints in Ho\v{r}ava-Lifshitz
gravity the problem with its closure.
In fact, after some algebra we find
following result
\begin{eqnarray}
& &\pb{\bT_T(f),\bT_T(g)}= \nonumber \\
&=&
\int d\sigma (f\parts g-\parts f g)
\left(\frac{2\pi\alpha'}{N^2\omega}F'
(p_t-N^kp_k)N^i h_{ij}\parts x^j+\frac{1}{\omega}
 F'^2(\dots)p_i\parts x^i\right) \ . \nonumber \\
\end{eqnarray}
This  result  shows
 that the algebra of constraints
$\bT_T$ is not closed. Then  as in case
of Ho\v{r}ava-Lifshitz gravity we see
that the algebra closes when we impose
 \emph{the projectability
condition} on the lapse function
$n_\tau$
\begin{equation}
n_\tau=n(\tau) \ .
\end{equation}
Then it is easy to see that the algebra
of constraints of LBS theory is closed
and takes the form
\begin{eqnarray}
\pb{\bT_S(\xi),\bT_S(\eta)}&=&
 \bT_S(\xi
\parts \eta-\parts \xi \eta) \ .
\nonumber \\
\pb{\bT_S(\xi),\bT_T(f)}&=&0 \ ,
\nonumber \\
 \pb{\bT_T(g),\bT_T(g)}&=& 0
\ , \nonumber \\
\end{eqnarray}
where $\bT_T(f)$ is defined
\begin{equation}
\bT_T=f(\tau)\int d\sigma
\mH_\tau(\sigma) \ .
\end{equation}
Let us now summarize consequences of
the breaking of Lorentz invariance of
the target space-time for the
construction of the string action in
given background.
 As the first one we have to
demand that the spatial part of the
world-sheet metric is dynamical.
Secondly, the full world-sheet
diffeomorphism
 is replaced with the
world-sheet
 foliation preserving
diffeomorphism. Finally, the
requirement of the consistent
Hamiltonian treatment of the LBS theory
implies the necessity of the
projectability condition on the lapse
function.

\vskip 5mm

 \noindent {\bf
Acknowledgements:}

 This work   was
supported by the Czech Ministry of
Education under Contract No. MSM
0021622409. \vskip 5mm


\end{document}